\documentclass[fleqn,12pt]{wlscirep}
\usepackage{lineno}

\usepackage{setspace}
\usepackage[utf8]{inputenc}
\usepackage[T1]{fontenc}
\usepackage{hyperref}

\usepackage[utf8]{inputenc} 
\usepackage[T1]{fontenc}    
\usepackage{hyperref}       
\usepackage{url}            
\usepackage{booktabs}       
\usepackage{amsfonts}       
\usepackage{nicefrac}       
\usepackage{microtype}      
\usepackage{lipsum}		
\usepackage{graphicx}
\usepackage{doi}
\usepackage{subcaption}

\title{The Temporal Rich Club Phenomenon}

\author[1,2]{Nicola Pedreschi}
\author[2,3]{Demian Battaglia}
\author[1,4]{Alain Barrat}

\affil[1]{Aix Marseille Univ, Universit\'e de Toulon, CNRS, CPT, Turing Center for Living Systems, Marseille, France}
\affil[2]{Aix-Marseille Univ, Inserm, INS, Institut de Neurosciences des Systèmes, Turing Center for Living Systems, Marseille, France
}
\affil[3]{University of Strasbourg Institute for Advanced Studies (USIAS), Strasbourg, France
}
\affil[4]{Tokyo Tech World Research Hub Initiative (WRHI), Tokyo Institute of Technology, Tokyo, Japan}

\begin{document}

\begin{abstract}
	Identifying the hidden organizational principles and relevant structures of networks representing complex physical systems is fundamental to understand their properties. To this aim, uncovering the structures involving a network's  prominent nodes in a network is an effective approach. In temporal networks, the simultaneity of connections is crucial for temporally stable structures to arise. We thus propose here a novel measure to quantitatively investigate the tendency of well connected nodes to form simultaneous and stable structures in a temporal network. We refer to this tendency, when observed, as the "\emph{temporal rich club phenomenon}". We illustrate the interest of this concept by analyzing diverse data sets under this lens, and showing how it enables a new perspective on their temporal patterns, from the role of cohesive structures in relation to processes unfolding on top of the network to the study of specific moments of interest in the evolution of the network. 
\end{abstract}

\keywords{Complex Networks \and Temporal Networks \and Complex Systems \and Rich Club phenomenon}

\maketitle


\section{Introduction}

A wide range of natural, technological and social systems can be represented as networks of agents (nodes) and their interactions (edges) \cite{baro,Dorogovtsev:2003a,Barrat2008DynamicalPO}. Typical examples include communication systems \cite{web}, transportation infrastructures \cite{Barrat3747}, biological and ecological systems \cite{bio-laszlo,Maslov910,skefi}, brain networks \cite{Bullmore:2009iv} or social interactions \cite{small-worlds,face-to-face,high-res}. 
The network representation offers a common framework and common tools to analyse the structure of these  systems, link their structure and dynamics and investigate processes on top of them. In particular, a common challenge in the study of networks consists in identifying relevant structures,
and several complementary approaches have been put forward to characterize networked data sets and their more central elements.
For instance, hubs, single nodes with very large numbers of connections (degrees), are known to influence spreading processes \cite{baro,Barrat2008DynamicalPO}. 
A quantification of a core-periphery structure identifies a central core of well-connected nodes 
\cite{coreperiphery}. The k-core decomposition \cite{alvarezhamelin:hal-00012974} decomposes the network into subgraphs of increasing connectedness, 
with correspondingly increasing influence in spreading processes \cite{influ-spreaders}. The rich-club coefficient quantifies whether the nodes with large numbers 
of neighbors (the hubs) 
tend to form more tightly interconnected groups 
\cite{Zhou2004rich,colizza,weighted,SerranoMA,across}
that can, for instance, share the control of resources in social and collaboration networks \cite{weighted}, 
or shape the routing and integration of communication in brain networks \cite{vandenHeuvel15775,Towlson6380, Nigam:2016fm}.

While all these approaches are effective for static networks, an increasing number of data sets include temporal information about edges, which can 
appear and disappear on different time scales: static networks are often only aggregated representations of the resulting temporal networks
\cite{Holme:2012,holme2015modern,holme:2019}, in which the information about the temporality of interactions has been lost. Thus, any structure
found in a static network obtained by temporal aggregation of data could in fact be formed by edges that were active at unrelated times. 
To investigate structures in temporal networks, it is thus crucial to take into account 
the complex temporal properties of the data.
For instance, various types of hubs can be defined, and a given node can be central during a certain period and peripheral in the next one \cite{pedreschi}; Network modular structures 
can evolve (which can e.g. be a resource for cognitive processing \cite{Braun:2015hj});
Processes can only take causal, time respecting paths among the elements of a network \cite{topo_dynamics,scholtes2015higher}; 
Concurrency, i.e., the 
simultaneity of connections of a given node with others, is key in epidemic propagation processes 
 \cite{masuda};
 Temporal motifs are defined as the 
 repetition of the connections in a small temporal subgraph in a given order
 \cite{kovanen2011temporal};
 Well connected structures such as cores are not static but are defined on specific time-intervals \cite{galimberti2018mining,span-cores}.

Overall, structures and hierarchies in temporal networks need to be defined and investigated taking into account (i) the temporality and simultaneity of the interactions forming
the structure,  (ii) the time-span on which the structure exists. Here, we propose a new way to investigate the cohesion of increasingly central nodes in a temporal network, namely,
the {\em temporal rich club coefficient}: given a temporal network, 
our aim is to quantify whether nodes who interact with increasing numbers of other nodes (i.e., with increasing degree in the aggregate network) 
tend also to interact {\em with each other simultaneously and in a stable way} (i.e., during a certain time period). 
We thus first define the \emph{$\Delta-$cohesion} of a group of nodes at each time $t$, as the density of links persistently connecting the nodes in the group 
during a time interval of length $\Delta$ starting at $t$. We then consider groups of nodes of increasing degree in the aggregated network, and measure
the maximum value of their $\Delta-$cohesion over time: this quantifies whether these groups
are tightly {\em and simultaneously} interconnected at least once {\em for a certain duration $\Delta$}. 
Moreover, and as in the case of the static rich club coefficient \cite{colizza}, a natural question is whether these simultaneous connections could 
exist just by chance, so that we compare the result with adequate null models for temporal networks \cite{gauvin2020randomized}.
To show the broad interest of this new analysis tool for temporal networks, we consider empirical temporal networks
representing very different systems: an air transportation infrastructure, a face-to-face interaction network in a social 
context, and a neuronal assembly, i.e., a network of neurons exchanging and integrating information. In each case, we compute the temporal rich-club coefficient
for the data and several null models, and highlight how it unveils interesting properties of the data. We show in particular how
static and temporal rich clubs are independent phenomena, how a temporal rich club impacts spreading processes, and how a temporal network undergoing
successive states \cite{state_sequences} can present a distinct temporal rich club in each state.
Our findings suggest that the temporal rich club coefficient provides a new tool in the complex analysis of temporal networks, shedding 
light on the role and connections of their most prominent elements and providing additional relevant information on the different periods of interest of the network.

\label{sec:introduction}

\section{Results}
\label{sec:results}

\subsection{The temporal rich club} 

\begin{figure}[h!]
    \centering
    \includegraphics[width=\linewidth]{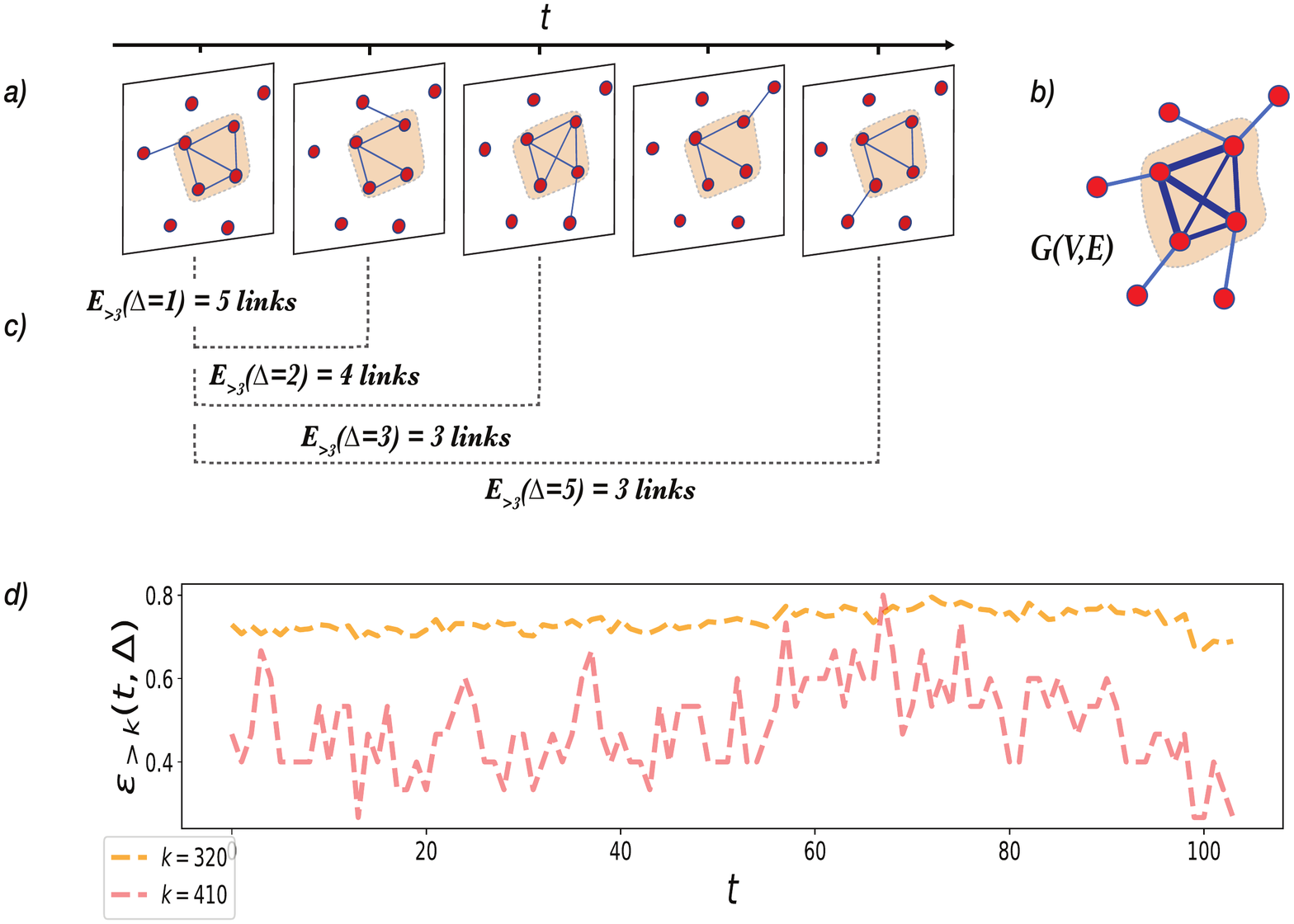}
    \caption{\textbf{a)} 
    Schematic representation of a temporal network as a sequence of instantaneous snapshots where nodes are connected by temporal edges. 
    \textbf{b)} Time aggregated graph $G(V,E)$, where the weight of an edge corresponds to the number of occurrences of the corresponding temporal edge.
    The  set $S_{>3}$ of nodes of degree larger than $3$ in the aggregate graph $G$ and its induced subgraph are included in the orange shaded area.
    \textbf{c)} Maximal number of edges among the nodes of $S_{>3}$ that are simultaneously stable over a duration $\Delta$, $E_{>3}(\Delta)$, for different values of $\Delta$. 
    \textbf{d)} Two examples of time series of the $\Delta-$cohesion $\epsilon_{>k}(t,\Delta)$ computed for the U.S. Air Transportation Temporal Network, 
    with $\Delta=1$; for $k=320$ (orange dotted line),  $\epsilon_{>320}(t,\Delta=1)$ remains persistently large, corresponding to a 
    \emph{stable} temporal rich club, while for $k=410$ the cohesion values fluctuate strongly, suggesting the existence of a \emph{transient} temporal rich club.}
    \label{fig:fig0}
\end{figure}

We consider a temporal network in discrete time on a time interval $[1,T]$ (Figure \ref{fig:fig0}.a): a temporal network can be represented as a series of instantaneous 
snapshots of the network at each time stamp. We denote by \emph{temporal edges} the interactions between pairs of nodes in each snapshot. 
The temporal aggregation over $[1,T]$ yields a static (aggregated) network 
 $G=(V,E)$ with set of nodes $V$ and set of edges $E$ (Figure \ref{fig:fig0}.b), 
 in which an edge is drawn between two nodes $i$ and $j$ if they have at least shared one temporal edge, with a
 weight $w_{ij}$  given by the number of temporal edges between $i$ and $j$. The degree $k$ of a node
 in $G$ is the number of distinct other nodes with which it has interacted at least once in $[1,T]$, and its strength $s$ the total number of temporal edges it has participated to.
 
 As stated above, our goal is to quantify a temporal rich club effect, corresponding to the fact that nodes of increasing degree in $G$ tend to be more connected
 than by chance {\em simultaneously and for a certain duration.} We first remind that the rich club coefficient was defined 
 for a static network as the density of edges in the subset $S_{>k}$ of the $N_{>k}$ nodes with 
 degree larger than $k$ \cite{Zhou2004rich,colizza}:
$\phi(k)=\frac{2E_{>k}}{N_{>k}(N_{>k}-1)} $,
where $E_{>k}$ is the number of edges connecting the $N_{>k}$ nodes. An increasing $\phi(k)$ indicates that nodes of larger degree tend to form increasingly connected
groups of nodes ("rich club effect"). However, such effect can be present even in random networks \cite{colizza}, so that 
the rich club ordering is detected by comparing $\phi(k)$ with the value 
obtained for a random network with the same degree sequence as the original one, $\phi_{ran}(k)$, i.e., by studying the ratio
$$
\rho(k)=\frac{\phi(k)}{\phi_{ran}(k)} \ .
$$
$\rho(k) > 1$ indicates indeed that the nodes with degree larger than $k$ are more connected than by chance.

Here, to take into account temporality,  we first define at each time $t$ the $\Delta-$cohesion
$\epsilon_{>k}(t,\Delta)$ as the number of ties between nodes of $S_{>k}$ that remain stable over the time interval
$[t,t+\Delta[$, normalized by the maximal possible value $N_{>k} (N_{>k}-1)/2$,
We then define the temporal rich club coefficient as the maximal cohesion observed in the temporal network over time:
$$
M(k,\Delta) \equiv \max_t \epsilon_{>k}(t,\Delta) \ .
$$
In other terms, $M(k,\Delta)$ is the maximal density of temporal edges observed in a stable way for a duration $\Delta$ among nodes of aggregated degree
larger than $k$. While, by definition, $M(k,\Delta)$ is non-increasing as a function of $\Delta$, a $M(k,\Delta)$ increasing with $k$ denotes that the most connected nodes
tend as well to be increasingly 
connected with each other in a simultaneous way for a duration at least $\Delta$. 
However, such simultaneity might also be observed by chance. To detect a temporal rich club effect, one needs therefore to compare $M(k,\Delta)$ with the value
$M_{ran}(k,\Delta)$
obtained in a suitable null model of the temporal network: 
$\mu(k,\Delta) \equiv M(k,\Delta) / M_{ran}(k,\Delta) > 1$ indicates that the nodes of degree larger than $k$ are more connected simultaneously on at least one time interval of duration $\Delta$ than 
expected by chance. Although there is a large variety of null models for temporal networks  \cite{gauvin2020randomized},
we focus here on randomization procedures that preserve the overall activity timeline of the temporal network (number of temporal edges at each time) as well
as the degree of each node in the aggregated graph (see Methods). In particular we will focus in the main text on a randomization that fully preserves $G$ (and thus its potential
static rich club), and present in the Supplementary Information (SI) the results obtained with two other randomizations.

Furthermore, as $M(k,\Delta)$ is defined as a maximum over time, it is also relevant to study the time evolution of the $\Delta-$cohesion $\epsilon_{>k}(t,\Delta)$,
in order to find the moments of highest simultaneous connectivity of $S_{>k}$, and to check whether this cohesion is stable or fluctuates strongly.
This allows for instance to distinguish between {\em stable or recurrent} and {\em transient} rich club effects: in the former case, $\epsilon_{>k}(t,\Delta)$ reaches
its maximum $M(k,\Delta)$ repeatedly, or remains close to it, while in the latter, $M(k,\Delta)$ is reached only once or only at specific moments.


\subsection{Static vs. temporal rich clubs}

\label{sec:air}

We first apply our measure on a data set describing the U.S. air transportation infrastructure from 2012 to 2020, 
with temporal resolution of one month, for $105$ snapshots (see Methods): 
in this temporal network, the $N=1920$ nodes represent airports and a temporal edge in one snapshot represents the 
existence of a direct connection in the corresponding month. 
The average number of temporal edges in a snapshot is $6126$ and, in the aggregated
network, the average degree is $44$, with degrees ranging from $1$ to $498$.

\begin{figure}[thb!]
	\centering
\includegraphics[width=\linewidth]{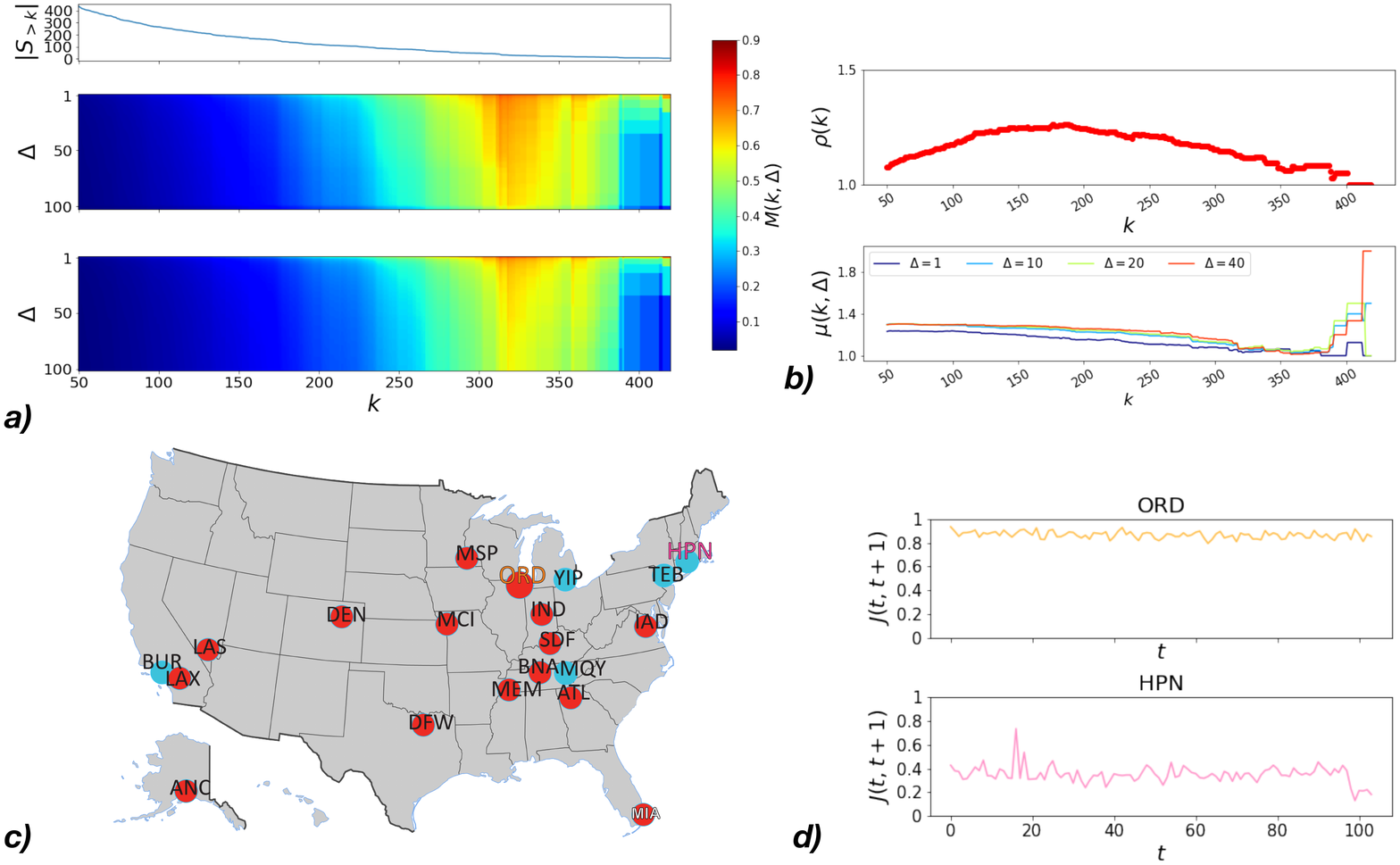}
	\caption{\textbf{U.S. air transportation temporal network.}
	\textbf{a)} (top) Size $N_{>k}=|S_{>k}|$ of the sub-network of nodes of aggregate degree larger than $k$ as a function of $k$; (middle) 
	temporal rich club coefficient $M(k,\Delta)$ as a color plot as a function of $k$  and $\Delta$, for the U.S. air transportation temporal network; 
	(bottom) $M_{ran}(k,\Delta)$ obtained for a randomized version of the temporal network that  preserves the activity timeline and the structure of the aggregated network.	
	\textbf{b)} (top) Static rich club coefficient $\rho(k)$  of the aggregated graph, as a function of the aggregate degree $k$;  $\rho(k)>1$ indicates
	that a rich club ordering is present \cite{colizza}, i.e., that the set of nodes $S_{>k}$ has more connections than expected by chance;
	(bottom) ratio $\mu(k,\Delta)$ between $M(k,\Delta)$ and $M_{ran}(k,\Delta)$ as a function of $k$ for specific values of $\Delta$. 
	$\mu(k,\Delta) > 1$ indicates that a temporal rich club ordering is present, i.e., that the interactions within $S_{>k}$ are more simultaneous than expected by chance.
	\textbf{c)} Geographic locations of the $20$ airports with largest aggregate degree ($S_{>350}$); 
	airports that are also in the group of $20$ nodes with highest strength ($s>10,000$, i.e., at least about $100$ different connections each month on average) in the aggregated network are depicted in red, 
	whereas the light blue nodes have low strength.  
	\textbf{d)} Jaccard index of the neighborhood of a node between times $t$ and $t+1$ as a function of time, computed for O'Hare International Airport (ORD, top), 
	and Westcherster County Aiport (HPN, bottom): both airports are in the top $20$ nodes for aggregate degree,
	 yet ORD is also in the set of $20$ nodes with largest aggregate strength, whereas HPN is not.}
	\label{fig:fig1}
\end{figure}

Figure \ref{fig:fig1}.a shows the $k-\Delta$ diagram of the temporal rich club coefficient $M(k,\Delta)$ as a color plot (the size of $S_{>k}$ being shown on top).
At fixed $k$, $M(k,\Delta)$ decreases as $\Delta$ increases (by definition, as larger $\Delta$ is a stronger requirement in terms of stability of temporal edges).
At fixed $\Delta$, $M(k,\Delta)$ is small for small and intermediate $k$, and decreases rapidly as $\Delta$ increases: 
many small airports have fluctuating activity, sometimes seasonal, so that many temporal edges involving these airports are not very stable \cite{Gautreau8847}, 
leading to a small cohesion at the global level. The maximal cohesion
however increases with $k$: airports with more connections tend also to be more interconnected and with increasingly stable connections (as found also in  \cite{Gautreau8847}).
$M(k,\Delta)$ reaches very large values around $k\sim 315$, even at large $\Delta$, indicating a stable and very cohesive structure. In fact, 
most of the $31$ airports in $S_{>315}$ are hubs of the U.S. air transportation system, which are  largely interconnected with very stable (and simultaneous) connections. 
For higher values of $k$, $M(k,\Delta)$ decreases again, especially at large $\Delta$, with a final increase close to the maximum possible value of $k$ (such that $|S_{>k}| \ge 2$).
This pattern indicates that, when restricting to $k > 380-390$, the interconnections of the nodes of $S_{>k}$ become actually less simultaneous and stable
than in $S_{>315}$: this indicates that some airports with degree larger than $380-390$ have actually {\em less} stable connections than others
with degree $315 < k < 380$, i.e., that some of the airports with very large aggregated degree have fluctuating connections. This is also clear from 
the timelines of $\Delta-$cohesion shown in Figure \ref{fig:fig0}.d for $k=320$ and $k=410$, with lower and more fluctuating values for $k=410$.

We further investigate this point in Figure \ref{fig:fig1}.c-d: Figure \ref{fig:fig1}.c shows the $20$ airports with largest aggregated degree, 
i.e., number of distinct other airports with which they share a direct connection
(degree values ranging from $350$ to $498$). 
We highlight in red the airports that are as well among the
$20$ nodes with largest aggregated strength ($s > 10,000$),
and in light blue the others. While the red nodes are typically well-known hubs, we find among the nodes in light blue airports such as
Burbank-Hollywood (BUR), Teterboro Airport (TEB) and Westchester County Airport (HPN). It turns out these airports serve 
as reliever airports for hubs such as LAX (Los Angeles) and JFK (New York), respectively: they are therefore extremely well connected in the aggregated network but 
have fluctuating connections, depending on the needs of the neighbouring hubs. Figure \ref{fig:fig1}.d highlights the differences between the two types of nodes, i.e.
the "real" hubs and the reliever airports, by displaying the Jaccard index between the connections of 
O'Hare International Airport (ORD, top plot) and Westchester County Airport (HPN, bottom) in successive months. 
ORD ($k=421$) has a very stable neighborhood while HPN
(reliever airport for JFK), despite
having the largest aggregated degree value $k=498$,  undergoes changes of up to $80\%$ of its neighborhood from a month to the next.

Figure \ref{fig:fig1}.a (bottom) displays for comparison the maximal cohesion $M_{ran}(k,\Delta)$ 
for a randomized version of the data, with conserved activity timeline and aggregated network, obtained by reshuffling the timestamps of the temporal edges
(see Methods, and Supplementary Material for other randomizations). $M_{ran}(k,\Delta)$ shows similar patterns but smaller values than $M(k,\Delta)$ for all
$(k,\Delta)$, showing that a temporal rich club ordering is present: for any $S_{>k}$, the interactions tend to be more simultaneously cohesive than expected by chance.
This is the case even at very large $k$: even when the reliever airports lead to a smaller $M(k,\Delta)$, its value is still larger than by chance.

{Differences with chance expectations are further investigated} in Figure \ref{fig:fig1}.b, which also highlights that the static and temporal rich club orderings show different patterns.
The top plot of the figure displays the normalized static rich club coefficient $\rho(k)$ (see also  \cite{colizza}):  $\rho(k) > 1$ indicates the presence of
a static rich club ordering, which becomes stronger as $k$ increases from $50$ to $\sim 250$. 
At large aggregated degree ($k \gtrsim 250$), $\rho(k)$ decreases, indicating that the density of links among the hubs tends to be 
closer to the one of a null model: these hubs have such large degree that, even in a null model, they tend to be largely interconnected.
The bottom plot of  Figure \ref{fig:fig1}.b shows that the ratio $\mu(k,\Delta)$ vs. $k$ for various $\Delta$ exhibits a different trend: 
$\mu(k,\Delta)$ is above $1$ and almost constant over a large range of $k$ values, and decreases for $320  \lesssim k \lesssim 380$: 
in this range of $k$ values, $S_{>k}$ is a mix of hubs and reliever airports, with both very stable connections and others much less stable. The randomization 
by time stamp reshuffling does not perturb the most stable connections, so that $M$ and $M_{ran}$ are closer. Finally for the largest aggregated degree values, $\mu(k,\Delta)$
reaches again very large values, especially for large $\Delta$: here many of the remaining connections are to reliever airports, which are not necessarily
very stable nor simultaneous, yet much more so than by chance.

The analysis of the US air transportation network under the lens of the temporal rich club can thus shed light on the different roles of well-connected nodes, and highlights how
temporal and static rich clubs can co-exist albeit with different patterns.

\subsection{Temporal rich club and spreading processes}

\label{sec:school}

The second dataset we consider is a temporal network of face-to-face interactions between
$232$ students and $10$ teachers of a primary school in France: the temporal edges between two nodes at a specific time stamp correspond to the detection by wearable
sensors of a face-to-face
interaction between the two corresponding individuals at that time \cite{face-to-face,high-res}
 (see Methods). The original time resolution of the dataset is $20$s for two schooldays, and, in order to smoothen the short time noisy dynamics,
 we perform a temporal coarse-graining on successive time-windows of $5$ minutes. We also consider in the main text the first school day only, i.e., 
 a temporal network of $N=242$ nodes and duration $T=103$ time stamps (each representing a $5$-minutes time window). The maximal degree in the aggregated network is $k_{max}=98$. Results for the whole 2-days data set and for a finer temporal resolution are shown in the SI.
 
\begin{figure}[thb!]
	\centering
	\includegraphics[width=\linewidth]{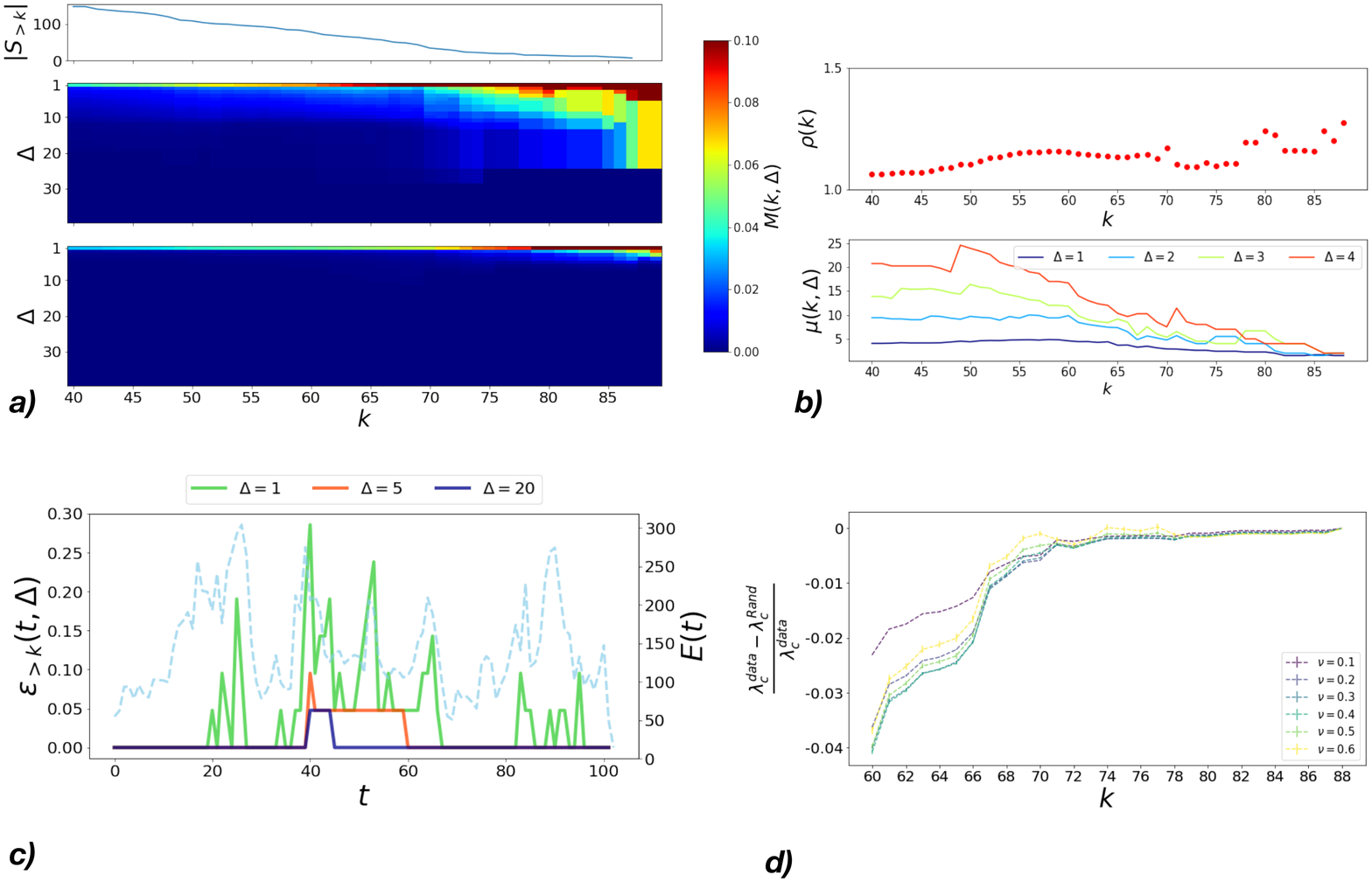}
	\caption{\textbf{Primary school temporal network.}
	 \textbf{a)}(top) Size $|S_{>k}|$ of the sub-network of nodes of aggregate degree larger than $k$ as a function of $k$ of the Primary School 
	 temporal network; 
	 (middle) Maximal cohesion $M(k,\Delta)$ as a function of $k$  and $\Delta$; 
	 (bottom) $M_{ran}(k,\Delta)$ diagram of the randomization preserving aggregate node statistics and overall activity timeline. 
	 \textbf{b)} 
	 (top) Static rich club coefficient $\rho(k)$ computed for the aggregated graph as a function of the aggregate degree $k$; 
	 (bottom) ratio $\mu(k,\Delta)$ between $M(k,\Delta)$, computed for the data, and $M_{ran}(k,\Delta)$, for different values of $\Delta$.
	 \textbf{c)} 
	 Instantaneous values of the cohesion $\epsilon_{>87}(t,\Delta)$ of $S_{>k}$ for various values of the temporal resolution $\Delta$, 
	 together with the instantaneous number of edges of the network $E(t)$ (dashed blue line).
	  \textbf{d)} 
	  Relative difference between the epidemic threshold $\lambda_c^{data}$, computed for the original dataset, and 
	  $\lambda_c^{rand}$, computed after the  randomization of the interactions between the nodes of $S_{>k}$ (see Methods).}
	\label{fig:fig2}
\end{figure}

Figure \ref{fig:fig2}.a displays the $k-\Delta$ diagrams of $M(k,\Delta)$ for the original temporal network (middle) and its randomized version ($M_{ran}(k,\Delta)$, bottom), with 
the size of $S_{>k}$ (top panel), as for Fig. \ref{fig:fig1}a.  At fixed $\Delta$, $M(k,\Delta)$ tends to increase with $k$; moreover,
 $M(k,\Delta)$ decreases more slowly with $\Delta$ when $k$ increases: nodes with higher degree in the aggregated network tends to be more tightly 
 interconnected, and in a more stable way.  For instance, the $7$ nodes of $S_{>87}$ 
keep a maximal cohesion $M(k,\Delta) \gtrsim 0.06$ up to $\Delta=25$. Notably, these structures disappear in the randomized version of the temporal network,
with much lower cohesion values on the whole $k-\Delta$ domain, indicating a temporal rich club ordering in the data.
This is confirmed in Figure \ref{fig:fig2}.b, which also highlights the differences between static and temporal rich clubs. The top plot displays
the normalized static rich club coefficient $\rho(k)$, which is larger than $1$ and tends to slightly increase, as with other social networks \cite{colizza}:
the children with a larger diversity of contacts (the degree in the aggregated network is the number of distinct other individuals contacted) tend also
to be more interconnected than expected by chance alone. For the temporal rich club coefficient, the ratio $\mu(k,\Delta)$ quantifies moreover the difference in simultaneous
interactions with respect to the randomized version: it is higher for larger $\Delta$, as stable simultaneous interactions are disrupted in the null model, remains stable on a broad
range of $k$ values, and tends to decrease at larger $k$.
This indicates that the nodes of the temporal network are connected in a much more simultaneous way than expected by chance, especially when considering stable interactions.

We investigate the dynamics of the temporal rich club in Figure \ref{fig:fig2}.c through the evolution of the instantaneous $\Delta-$cohesion
$\epsilon_{>k}(t,\Delta)$ of the $7$ nodes with aggregated degree  $k>87$ for several values of $\Delta$. 
We show also for reference the activity timeline of the network: the simultaneous cohesion of these nodes fluctuates strongly, is $0$ in many snapshots and reaches its maximum
in the periods of high overall activity (namely recess and lunch break \cite{high-res}), forming a transient but repeated temporal rich club. We have verified that these students actually 
belong to different school classes, which explains why the moments of highest cohesion of this group can only happen during the breaks.

The temporal network under scrutiny represents interactions among individuals, which can be the support of many processes, and in particular of the spread of information
or infectious diseases. It is thus relevant to investigate whether the temporal rich club ordering plays a role in the unfolding of such processes, as 
with other temporal structures \cite{span-cores}. We therefore consider the paradigmatic susceptible-infected-susceptible (SIS) model of spreading processes, 
in which nodes can be either susceptible (S) or infectious (I): a susceptible can become infectious upon contact with an infectious, with
probability $\lambda$ per time step; infectious individuals recover with probability $\nu$ at each time step and become susceptible again. We quantify
the interplay between the temporal network and the spread by the epidemic threshold  $\lambda_c$ at given $\nu$: it separates a phase at $\lambda < \lambda_c$ 
in which the epidemic dies out from a phase at $\lambda > \lambda_c$ where it reaches a non-zero fraction of the population. We compute the epidemic threshold,
using the method of  \cite{valdano}, in (i) the original data set ($\lambda_c^{data}$) and (ii) versions of the data set in which the temporal edges connecting the nodes in  
 $S_{>k}$ are randomized ($\lambda_c^{rand}$), thus disrupting their simultaneity (note however that such temporal randomization leaves the static rich club unaltered, see Methods). Figure \ref{fig:fig2}.d displays the relative difference between the two obtained values 
 as a function of $k$. This difference takes higher absolute values for lower values of $k$, which can be expected as the randomization affects then a larger number of temporal edges;
 most importantly,  $\lambda_c^{data}$ is systematically lower than $\lambda_c^{rand}$: this indicates that the spreading process is favoured by
 the temporal rich club of the data, i.e., by the stronger simultaneity of connections than in the randomized versions \cite{masuda}. The effect is also larger for larger $\nu$, i.e., for faster
 processes. Cohesive simultaneous structures of prominent nodes in a temporal network, as revealed by the temporal rich club ordering, can thus affect spreading dynamics unfolding on top of the network.

\subsection{State-specific temporal rich clubs}

\label{sec:neuro}

We finally investigate the temporal rich club patterns of a network of biological relevance, namely the 
time-resolved functional connectivity of $N=67$ neurons in the entorhinal cortex and hippocampus of an anesthetized rat. 
The nodes represent single neurons and the temporal edges correspond to a significant mutual information between the firing patterns
of pairs of neurons in a sliding window of 10 seconds \cite{pedreschi, Clawson:2019jq}. Successive time windows are shifted of $1$ second: this is the temporal resolution of the network,
which lasts $2284$ seconds.

\begin{figure}[thb!]
	\centering
	\includegraphics[width=\linewidth]{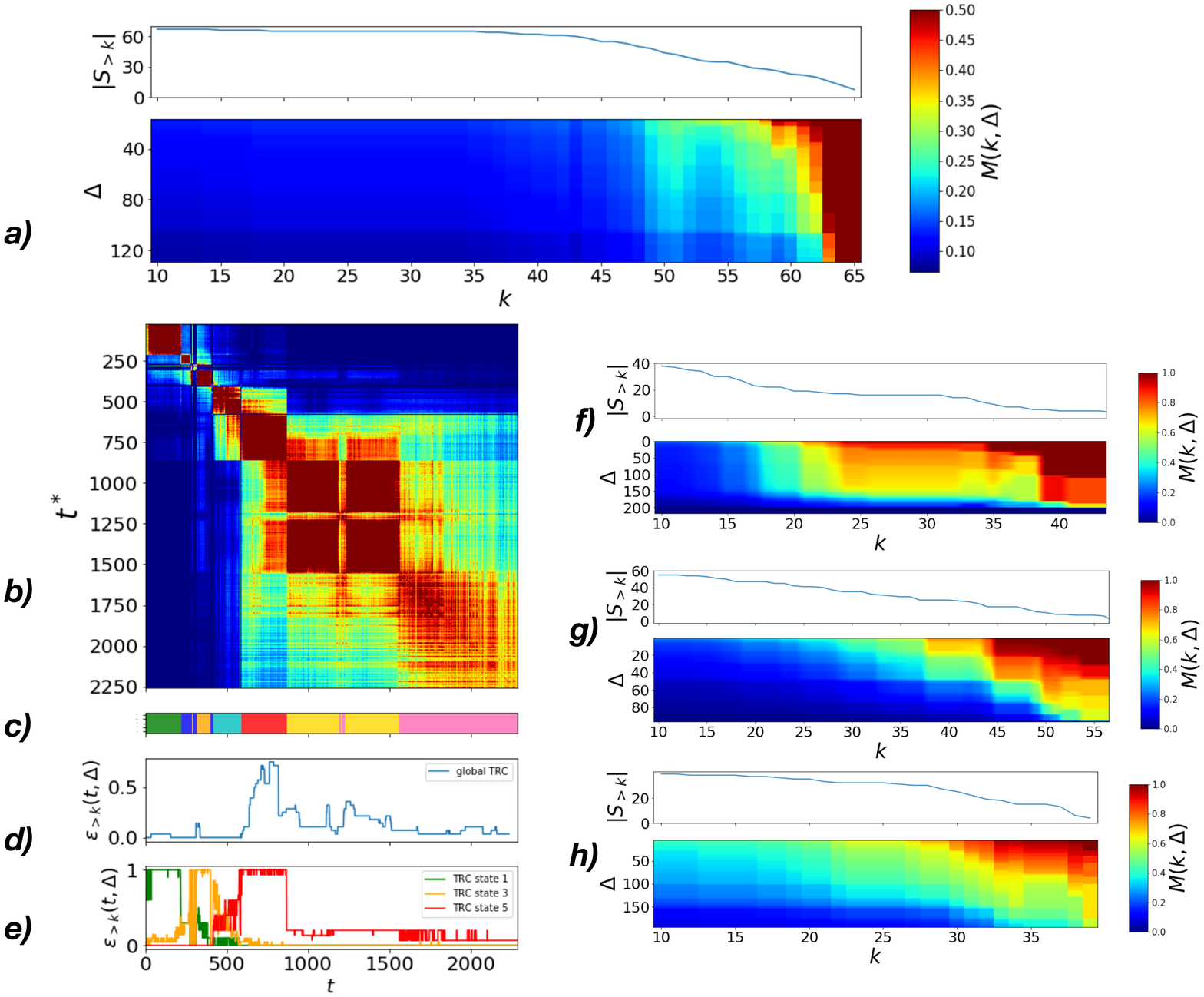}
	\caption{\textbf{Temporal network of information sharing neurons.} 
	\textbf{a)} (top) Size $|S_{>k}|$ of the sub-network of nodes of aggregate degree larger than $k$ as a function of $k$; 
	(bottom) Maximal cohesion $M(k,\Delta)$ as a function of $k$  and $\Delta$. 
	\textbf{b)} Temporal network similarity matrix: the $(t,t^*)$ matrix entry is given by the similarity
	between the  instantaneous snapshots of the network at times $t$ and $t^*$. The red blocks around the diagonal indicate periods in which the network remains similar to itself,
	i.e., "states" of the network \cite{masuda,pedreschi}.
	\textbf{c)} Timeline of the states of the network, represented as a colored barcode (each color represents a different state), as extracted by clustering of the 
	similarity matrix in \cite{pedreschi}. 
	\textbf{d)} Instantaneous cohesion $\epsilon_{>65}(t,\Delta)$ of the $N_{>65}=8$ nodes of aggregate degree larger than $k=65$. 
	\textbf{e)} 
	Instantaneous cohesion $\epsilon_{>k}(t,\Delta)$ of the nodes with highest degree in the aggregate graphs of states $1$, $3$ and $5$, as a function of time
	during the whole recording (respective largest degree values: 
	$45$, $57$ and $40$, same color code as in panel c)); 
	the sets have sizes $|S^1_{>44}|=5$ (cohesion in green), $|S^3_{>56}|=7$ (orange) and $|S^5_{>39}|=7$ (red); 
	$S^1_{>44}$ has $1$ node in common with $S^3_{>56}$, 
	and $S^5_{>39}$ has no node in common with 
	$S^1_{>44}$ nor with $S^3_{>56}$. 
	\textbf{f-h)} For each of the states $1$, $3$ and $5$, size
	 $|S_{>k}|$ as a function of $k$ in the aggregate network of the state, and color-plot of the 
	 temporal rich club coefficient $M^s(k,\Delta)$ for the temporal network restricted to the same state.}
	\label{fig:fig3}
\end{figure}

We first note that the aggregated network is very dense: the average degree is $\langle k\rangle=54$  (whereas the minimal value of $k$ is $k_{min}=14$) and
the maximal degree is equal to $N-1=66$. In such a dense network, the static rich club ordering cannot be assessed as randomization of the links of the high degree
nodes cannot be achieved. Taking into account temporality reveals a much richer picture. 
Figure \ref{fig:fig3}.a shows that the temporal rich club coefficient $M(k,\Delta)$ increases with $k$ for each value of $\Delta$, and that
higher values of $k$ are needed to reach a given cohesion when $\Delta$ increases: groups of nodes with increasing aggregated degree are simultaneously
interconnected for increasing durations. The group of $8$ neurons with largest degree 
(which are each connected at least once over the temporal network duration to each of the other nodes) 
are in particular very strongly interconnected in a simultaneous way, 
with $M(k,\Delta)\geq 0.5$ up to $\Delta=140$. We show in the SI that $M_{ran}(k,\Delta)$ takes much smaller values and do not exhibit any relevant structure, indicating the existence of a temporal rich club in this data set.

As investigated in \cite{pedreschi}, the temporal network of functional connectivity actually goes through several "states", found through the hierarchical clustering
of the network similarity matrix shown in Figure \ref{fig:fig3}.b \cite{state_sequences,pedreschi}: each element $(t,t^*)$ of this matrix gives the 
similarity between the snapshots of the network at times $t$ and $t^*$, and periods of stability of the network ("states") are found as periods of large similarity values
(red blocks along the diagonal). The timeline of successive states is shown in Figure \ref{fig:fig3}.c, and Figure \ref{fig:fig3}.d shows 
the instantaneous cohesion $\epsilon_{>65}(t,\Delta=1)$
of the nodes with highest degree in the network aggregated on the whole recording: the cohesion among these nodes changes strongly from one state to another, and
actually reach very large values only during one specific state. We thus investigate separately these different states, computing 
an aggregated network $G^s$ for each state $s$ by aggregating the temporal edges in the snapshots belonging to $s$, and defining
$S_{>k}^s$ as the set of nodes with degree larger than $s$ in $G^s$: nodes are not similarly active in each state and 
have thus different degrees in the different $G^s$. This leads us to measure the state-specific temporal rich club coefficients $M^s(k,\Delta)$. 
Figure \ref{fig:fig3}.f-h show that the corresponding  $k-\Delta$ diagrams for states $1$, $3$ and $5$ have similar but distinct patterns, with in each case
more stable simultaneously interconnected sets of nodes as $k$ increases, i.e., a temporal rich club structure (we show in the SI that the randomized data sets yield much lower values of $M$).

Furthermore, Figure \ref{fig:fig3}.e displays the $\Delta-$cohesion over time of the sets of nodes with largest degree in each of these states: notably,
this instantaneous cohesion is maximal (and reaches the maximal possible value $\epsilon_{>k}(t,\Delta)=1$) precisely in the time stamps of the
corresponding state. Note that the sizes of $|S^1_{>44}|=5$, $|S^3_{>56}|=7$, $|S^5_{>39}|=6$ in the three states are comparable, but that 
the nodes belonging to these three sets are mostly different: of the nodes in $S^1_{>44}$ only one is also in $S^3_{>56}$, and $S^5_{>39}$ has an empty intersection with the other sets.

Overall, the analysis of this temporal network highlights how a temporal rich club phenomenon 
can be present even when a static rich club cannot be identified. Moreover, it shows that 
distinct temporal rich clubs can be found when a temporal network goes through different states. 
Further investigation of the mutual relations of the state-wise temporal rich clubs could help shed light on the function of the different states of the system \cite{pedreschi}.

\section{Discussion}
\label{sec:Discussion}

In this paper we have defined a novel concept to investigate temporal networks and quantify the patterns of simultaneous interconnectedness
of nodes, namely the Temporal Rich Club. We have defined the temporal rich club coefficient as the maximal value of the 
density of links stable during at least a duration $\Delta$
between nodes having aggregated degree at least $k$. We compare the values obtained on an empirical data set need to those reached in 
randomized versions of the data, in which the activity timeline as well as the 
properties of the aggregated network (and potentially its whole structure) are preserved, in order to measure whether the 
simultaneity and stability of the connections of groups of nodes is higher than expected: a temporal rich club ordering corresponds
indeed to higher simultaneous cohesion than expected by chance. We note here a delicate point: 
many randomization procedures are possible for a temporal network \cite{gauvin2020randomized}, so the comparison with randomized data could be done in several ways. As the focus is on the simultaneity of connections, we have limited ourselves
to reshuffling procedures that maintain the network activity timeline and the aggregated properties of the nodes. Other reshuffling methods 
could however be considered.

In two of the data sets we have explored, both static and temporal rich club ordering were present. In general however, 
a static rich club in the aggregated graph could exist with or without a temporal rich club, as the links of the static network could correspond to interactions
occurring at different times. Vice-versa, interactions could be more simultaneous than expected by chance, with a temporal rich club ordering, 
without forming a static rich club. This is the case in the third data set explored, where the large density of the aggregated network makes the static rich club
concept irrelevant, while taking into account temporality reveals a more interesting picture with a temporal rich club ordering.
 
A limit of our analysis comes from the fact that we have considered 
the degree of nodes in the aggregate network as the reference for centrality in the aggregate network. 
A natural extension of this study would be to consider instead the strength of the nodes in the aggregate network, i.e., the number of temporal edges
to which they have participated during the span of the temporal network: in this case, the focus would be to investigate the simultaneity of the connections
within the set $S_{>s}$ of nodes having participated to more than $s$ temporal edges. As strength and degree are generally correlated, the results are expected to be similar,
but some significant and interesting differences might emerge, as in the example of the 
air transportation temporal network where the reliever airports have a very high degree but relatively low strength.

Overall, the temporal rich club perspective provides a new tool to study temporal networks and in particular to 
unveil the relevance of simultaneous interactions of increasingly connected nodes in 
processes unfolding on top of the temporal network: we have shown for instance that a temporal rich club pattern
favours spreading dynamics, similarly to other static or temporal cohesive structures \cite{influ-spreaders,span-cores,masuda}, 
suggesting to add such new measure to the repertoire of methods to study contagion processes in networks.
Moreover, we have shown how distinct temporal rich club patterns can be found when a temporal network evolves through different states,
and provide thus an additional way to characterize such states and, possibly, investigate their function. For instance, key processes in neural information processing, such as synaptic plasticity, are critically affected by the timing of neuronal interactions \cite{Markram:2012ik} and different temporal rich clubs in different states may thus enable flexible computations within a same circuit \cite{Clawson:2019jq}.
In conclusion, our work provides a new procedure to detect relevant temporal and structural patterns in a general temporal network,
enabling a new quantitative perspective on the temporal patterns of data sets coming from very different fields,
from highlighting the role of simultaneous connections between central nodes in spreading on a 
temporal network of social interactions to that of hubs in air transportation infrastructures or in neuronal assemblies.

\section{Methods}
\label{sec:methods}

\subsection{Data}

We consider three publicly available data sets. We have moreover gathered them at 

https://github.com/nicolaPedre/Temporal-Rich-Club/

\paragraph{Air Transportation Network.}
This data set represents the connections between US airports, with temporal resolution of one month, 
from January $2012$ to September $2020$,
for a total of $105$ time stamps. 
The $N=1920$ nodes of the temporal network represent the airports, and in each monthly snapshot a temporal  
edge is drawn between two nodes if there was at least one direct flight between the corresponding airports during that month. The degree of a node in the aggregated network is thus the number of other airports to which it has been connected directly once, and its strength is its total number of
temporal edges.
The data is publicly available on the website of the Bureau of Transportation Statistics (https://www.transtats.bts.gov/, "Air Carrier Statistics (From 41 Traffic) - U.S. Carriers" data base).

\paragraph{Face to face interactions.} 
This data set describes the face-to-face close proximity contacts between $232$ children and $10$ teachers in a Primary School of Lyon, France, during two days in 
2009, as collected by the SocioPatterns collaboration using wearable devices. The original data is publicly available 
from the SocioPatterns website (http://www.sociopatterns.org/datasets/primary-school-temporal-network-data/). 
The original data is a temporal network with temporal resolution of $20$s, where the nodes represent the individuals and each temporal edge corresponds to the detection
of a face-to-face contact between them \cite{high-res}. Here, we perform a temporal coarse-graining on successive time-windows of $5$ minutes to remove
short-time noise. The results in the main text correspond to the first day of data while the analysis performed on the whole data set can be found in the Supplementary Material,
as well as the results obtained with a coarse-graining on time-windows of $1$ minute.

\paragraph{Information sharing neurons.} 
This data set describes the functional connectivity between neurons in the hippocampus and medial entorhinal cortex of an anesthetized rat. 
The data was first presented in \cite{Clawson:2019jq} and further analysed in \cite{pedreschi}.
The network analysed here is made of $N=67$ nodes.
Each temporal edge represents a ``functional connection'', i.e. the existence of a significant mutual information between the firing patterns of the corresponding pair of neurons 
computed in a sliding window of $10$ seconds.  Overlapping sliding windows are considered, each being shifted of $1$ second with respect to the previous one.
The duration of the temporal network is $T=2284$ seconds. More details about the computation of time-resolved functional connectivity can be found in the original studies \cite{Clawson:2019jq, pedreschi}.

\begin{table}[h!]
	\caption{Some properties of the data sets}
	\centering
	\begin{tabular}{lllllll}
		\toprule
		Data      &  $N$ & $k_{min}$ & $k_{max}$ & $\langle k\rangle$  &  $\#$ temporal edges & time resolution, $T$\\
		\midrule
		U.S. Airways & $1,920$ & $1$&$498$  & $44$ & $1,286,616$  &  $t=1$ month, $T=105$     \\
		Primary School     & 242 & $1$ & $98$&$49$  &  $53,056$ &  $t=5$ minutes, $T=103$      \\
		Info. sharing network     & 67  &  $14$ & $66$ & $53$ &    $511,174$    & $t=1$s, $T=2,284$ \\
		\bottomrule
	\end{tabular}
	\label{tab:table}
\end{table}

\subsection{Temporal network randomization}

We consider a temporal network in discrete time $TN(V, \Gamma,T)$ as a set of nodes $V=\{i=1,2,3\dots N\}$ and a set of temporal 
edges $\Gamma=\{\gamma_1,\gamma_2,\dots\gamma_\Gamma\}$ where each temporal edge $\gamma_q=(i_q,j_q,t_q)$ 
represents an interaction between nodes $i_q$ and $j_q$ at time $t_q \in [0,T]$. 
An event or contact $(i,j,t,\tau)$ is moreover defined as an uninterrupted succession of temporal edges between nodes $i$ and $j$ starting at $t$ and lasting $\tau$ time steps, i.e., 
a series of temporal edges $(i,j,t),(i,j,t+1),\cdots,(i,j,t+\tau-1)$. Furthermore, the "activity" of the temporal network at time $t$ is simply the number of temporal
edges at $t$. The aggregated network $G=(V,E)$ is obtained by drawing an edge between all pairs of nodes that have interacted at least once during $[0,T]$.

A wide range of randomization procedures (null models) exist for temporal networks \cite{gauvin2020randomized}. Here, our focus is on the simultaneity and stability of interactions, which
define the existence of temporal rich club phenomena. As simultaneous interactions can occur simply by chance in periods of larger
activity, we will consider randomization procedures that preserve the temporal activity timeline, i.e., the number of temporal edges at each time step.
Moreover, in order to investigate the role of temporality, we need to consider procedures that keep either the whole structure of the aggregated network $G$,
or at least the degree of each node.

In the present work, we consider moreover two types of data randomization: either the randomization of the whole temporal network, or a randomization
involving only the subgraph induced by the set $S_{>k}$ of nodes of degree larger than $k$ in the aggregate graph. 
The former case allows us to compare the temporal rich club coefficient $M(k,\Delta)$ of the data with the values obtained for a null model, and thus to detect
whether a temporal rich club ordering is present.
The latter case is used in Section \ref{sec:school} to compare the epidemic threshold in the original data and in a partially reshuffled data set where the simultaneity
of the connections between the nodes of degree larger than $k$ (i.e., in $S_{>k}$) is disrupted while the rest of the connections are unchanged.

\paragraph{Randomized reference models.} 
We consider three ways to randomize temporal networks among the methods described in \cite{gauvin2020randomized}. In the main text
(Figures \ref{fig:fig1}.a, \ref{fig:fig2}.a and \ref{fig:fig3}.a,c), we show the resulting $M_{ran}(k,\Delta)$ for the first of these three null models, while the 
results for the other two are shown in the Supplementary Information.
\begin{itemize}

    \item \textit{Timestamps shuffling}: this reshuffling procedure, denoted $P[w,t]$ in \cite{gauvin2020randomized}, randomly 
    permutes the timestamps $t_q$ of all temporal edges while keeping the nodes indices $i_q$ and $j_q$ fixed. 
    This randomization therefore conserves the overall activity timeline of the network as well as the structure and weights of the edges in the aggregated graph $G$;
    
    \item \textit{Event shuffling, or Topology-constrained snapshot shuffling}, denoted $P[{\cal L}, p(t,\tau)]$ in \cite{gauvin2020randomized}: this procedure
    shuffles the contacts between existing links while keeping the contacts' starting time and duration, thus preserving contact duration statistics, global activity timeline and structure of the aggregated network.
    
    \item \textit{Degree-constrained link shuffling}, denoted $P(k, p_{\cal L} (\Theta) )$ in  \cite{gauvin2020randomized}: it 
    permutes the edges in the aggregated graph and associated timelines between all node pairs $(i,j)$ while keeping the aggregate degree
    $k$ of each node fixed. The randomization is typically implemented following the Maslov-Sneppen method on the aggregate graph $G$ \cite{Maslov910}.
    It preserves the global activity timeline and the aggregated degree of each node, as well as the distribution of edge weights, 
    but randomizes the structure of the aggregated network.
    
\end{itemize}
For each data set and each randomization procedure, we compute $100$ realisations of the randomized data set and compute the average $M_{ran}$ of the temporal rich club coefficient over these realisations.

\paragraph{$S_{>k}$ randomization.} 
We also consider a randomization restricted to the temporal edges joining the nodes of degree larger than $k$ in $G$. Specifically, we apply the timestamps shuffling
to these edges, while keeping all the other temporal edges fixed. The activity timelines of both the whole temporal network and of the nodes of $S_{>k}$ 
are thus preserved, as well as the structure of the aggregate edges of the subgraph induced by $S_{>k}$. However, 
the simultaneity of the connections between the nodes of $S_{>k}$ is disrupted.

\section*{Acknowledgements}

N.P. has received funding from the European Union’s Horizon 2020 research and innovation programme under the Marie Skłodowska-Curie grant agreement No. 713750. Also, the project has been carried out with the financial support of the Regional Council of Provence- Alpes-C\^ote d’Azur and with the financial support of the A*MIDEX (ANR-11-IDEX-0001-02), funded by the Investissements d’Avenir project funded by the French Government, managed by the French National Research Agency (ANR).

\noindent
D.B. was supported by the European Union Innovative Training Network “i- CONN” (H2020 ITN 859937)

\noindent
A.B. is partially supported by the Agence Nationale de la Recherche (ANR) project DATAREDUX (ANR-19-CE46-0008).

\bibliography{references}  

\newpage 
\setcounter{section}{0}
\setcounter{table}{0}
\setcounter{equation}{0}
\setcounter{figure}{0}
\renewcommand{\figurename}{Supplementary Figure}

\renewcommand{\thesection}{S\arabic{section}}
\renewcommand{\thetable}{S\arabic{table}}
\renewcommand{\thefigure}{S\arabic{figure}}

\section{US Airline data set randomization}
\begin{figure}[h!]
    \centering
    \includegraphics[width=\linewidth]{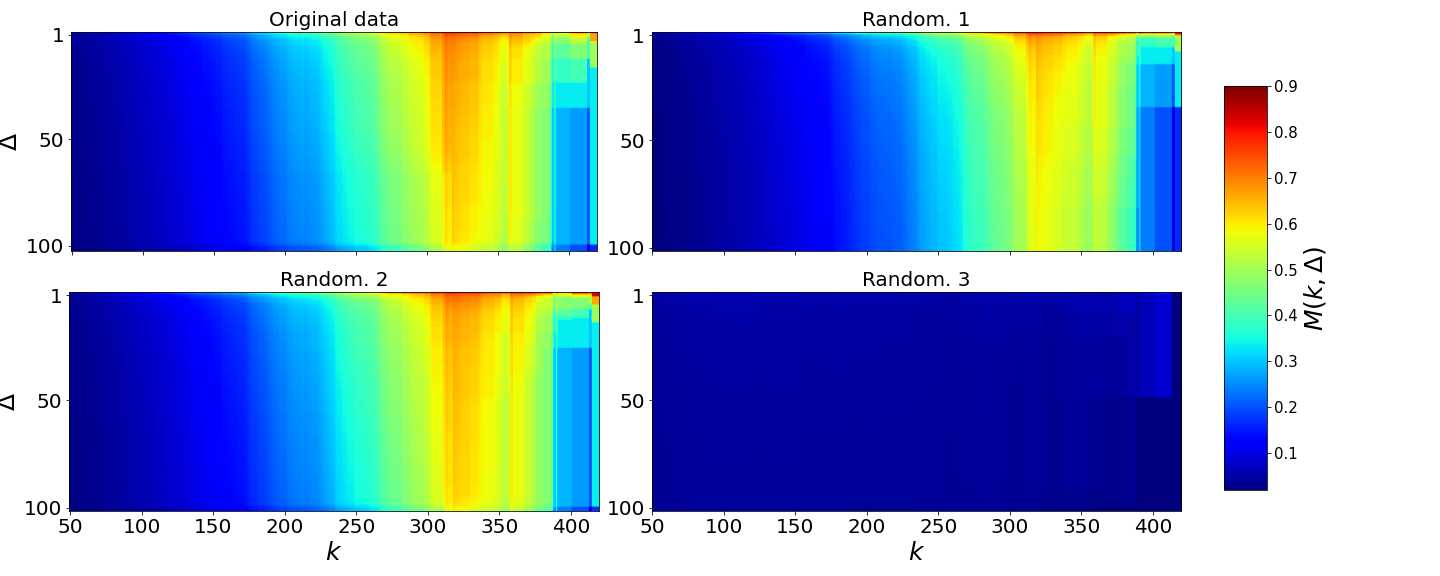}
    \caption{\textbf{Air transportation temporal network:} the four plots correspond to the $k-\Delta$ diagrams of the maximal cohesion $M(k,\Delta)$ of the original data set (top left, shown also in Figure 2.a), the \emph{Timestamps reshuffling} (top right, shown also in Figure 2.a) randomization of the temporal network and the \emph{Event} (bottom left) and \emph{Degree-constrained link} (bottom right) shufflings as described in Methods.}
    \label{fig:SI_1}
\end{figure}

\newpage

\section{Primary school data set reshuffling and different time resolution}
\begin{figure}[h!]
    \centering
    \includegraphics[width=\linewidth]{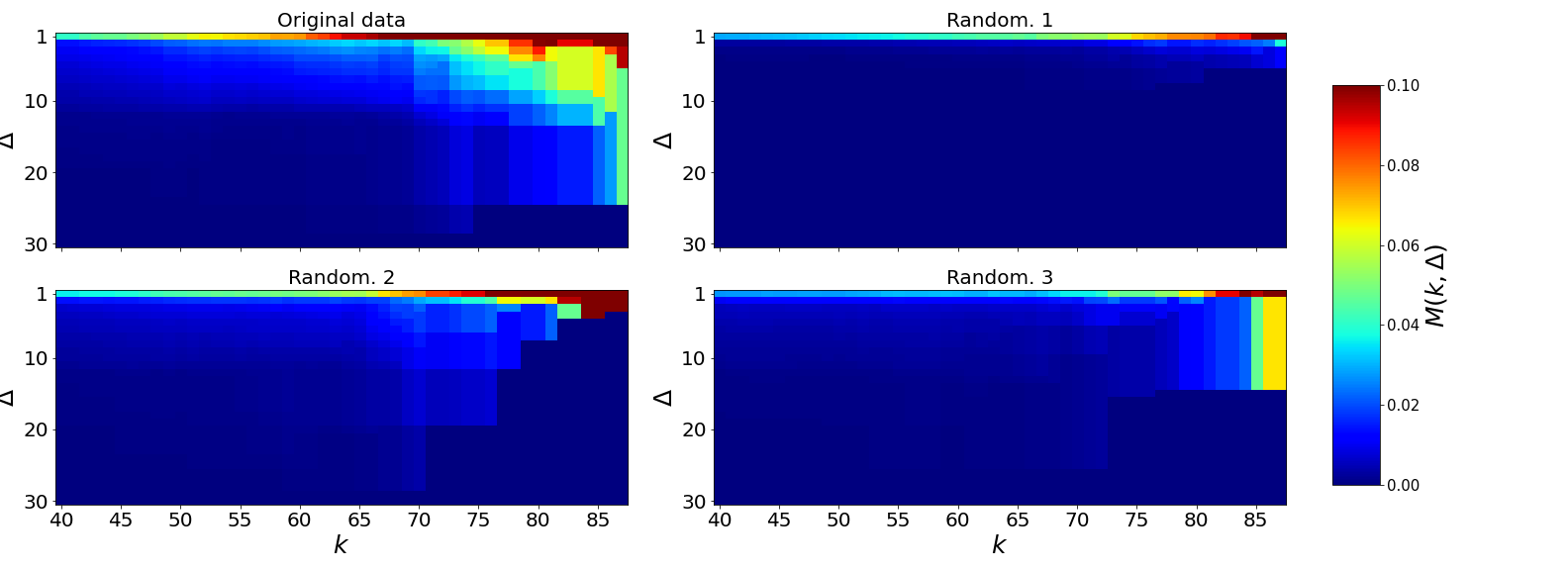}
    \caption{\textbf{Primary School temporal network:} the four plots correspond to the $k-\Delta$ diagrams of the maximal cohesion $M(k,\Delta)$ of the original data set (top left, shown also in Figure 3.a), the \emph{Timestamps reshuffling} (top right, shown also in Figure 3.a) randomization of the temporal network and the \emph{Event} (bottom left) and \emph{Degree-constrained link} (bottom right) shufflings as described in Methods.}
    \label{fig:SI_2}
\end{figure}

\begin{figure}[h!]
    \centering
    \includegraphics[width=\linewidth]{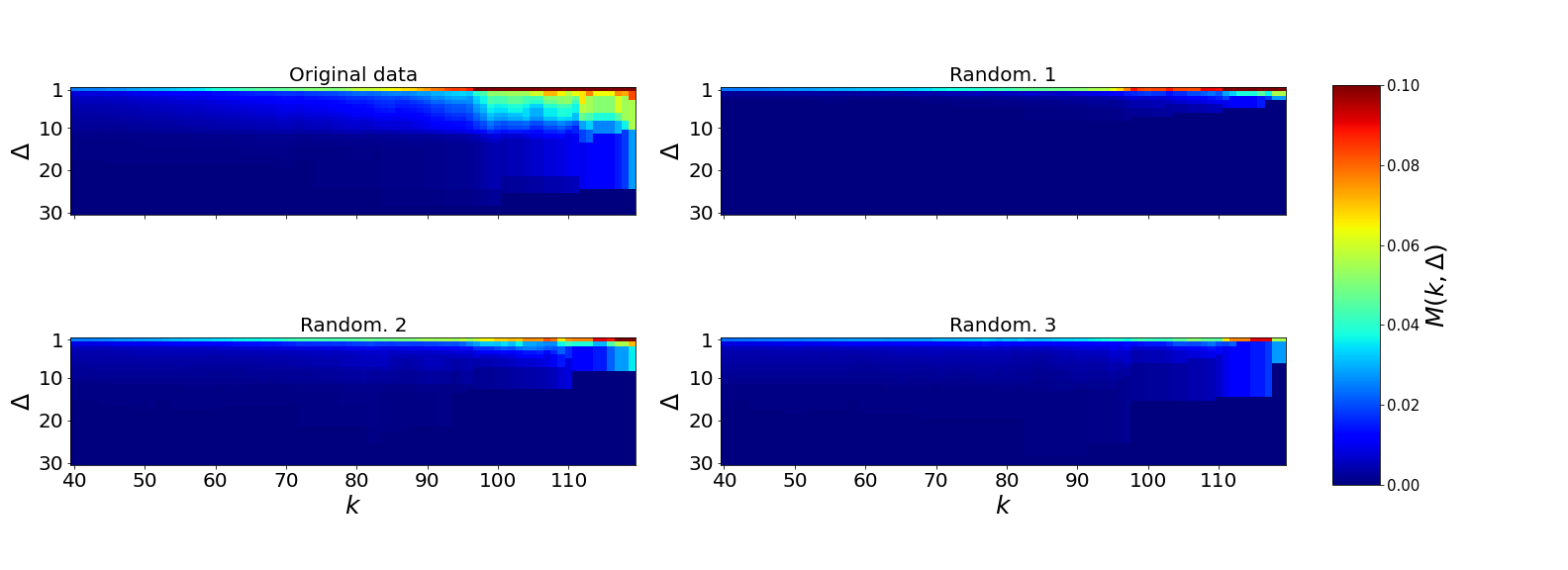}
    \caption{\textbf{Primary School temporal network, $2$ days:} the top left plot corresponds to the $k-\Delta$ diagram of the maximal cohesion $M(k,\Delta)$ of the original data set computed over two school days. The other three plots are the $k-\Delta$ diagram of the maximal cohesion $M(k,\Delta)$ of the \emph{Timestamps reshuffling} (top right) randomization of the temporal network and the \emph{Event} (bottom left) and \emph{Degree-constrained link} (bottom right) shufflings as described in Methods.}
    \label{fig:SI_3}
\end{figure}

\begin{figure}[h!]
    \centering
    \includegraphics[width=\linewidth]{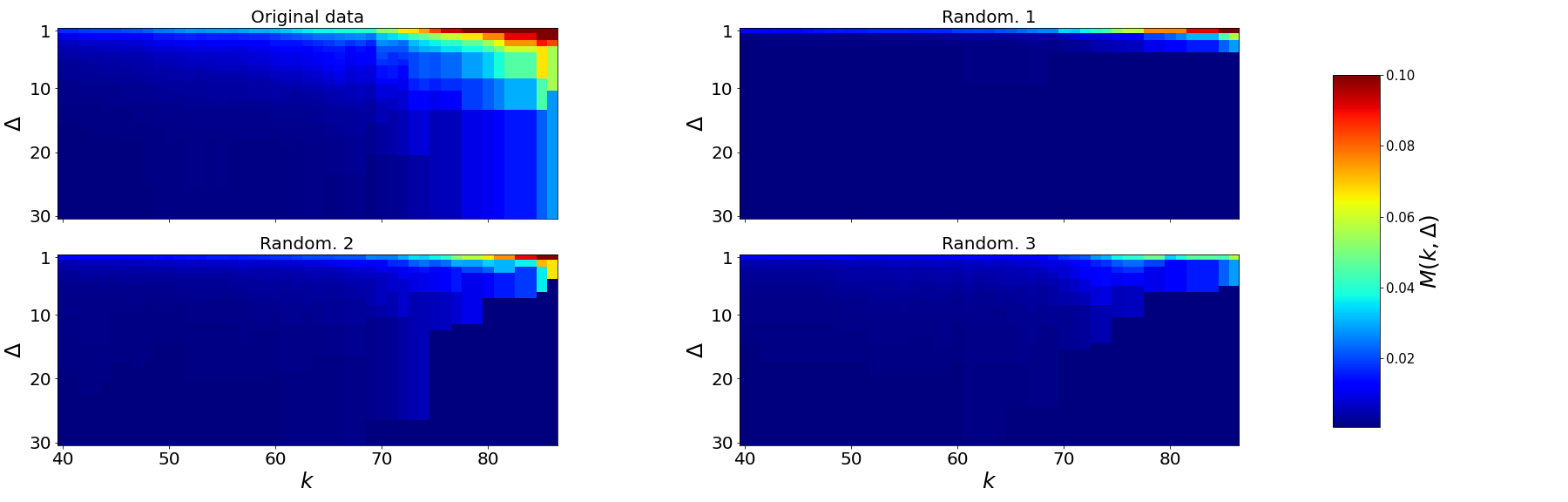}
    \caption{\textbf{Primary School temporal network, $1st$ day, $1$ minute time resolution:} the top left plot corresponds to the $k-\Delta$ diagram of the maximal cohesion $M(k,\Delta)$ of the original data set computed over one school day, with a partial time aggregation of $1$ minute ($3$ successive $20$s-snapshots). The other three plots are the $k-\Delta$ diagram of the maximal cohesion $M(k,\Delta)$ of the \emph{Timestamps reshuffling} (top right) randomization of the temporal network and the \emph{Event} (bottom left) and \emph{Degree-constrained link} (bottom right) shufflings as described in Methods.}
    \label{fig:SI_4}
\end{figure}

\clearpage
\newpage

\section{Neuronal assembly reshuffling, whole recording and state-wise}
\begin{figure}[h!]
    \centering
    \includegraphics[width=\linewidth]{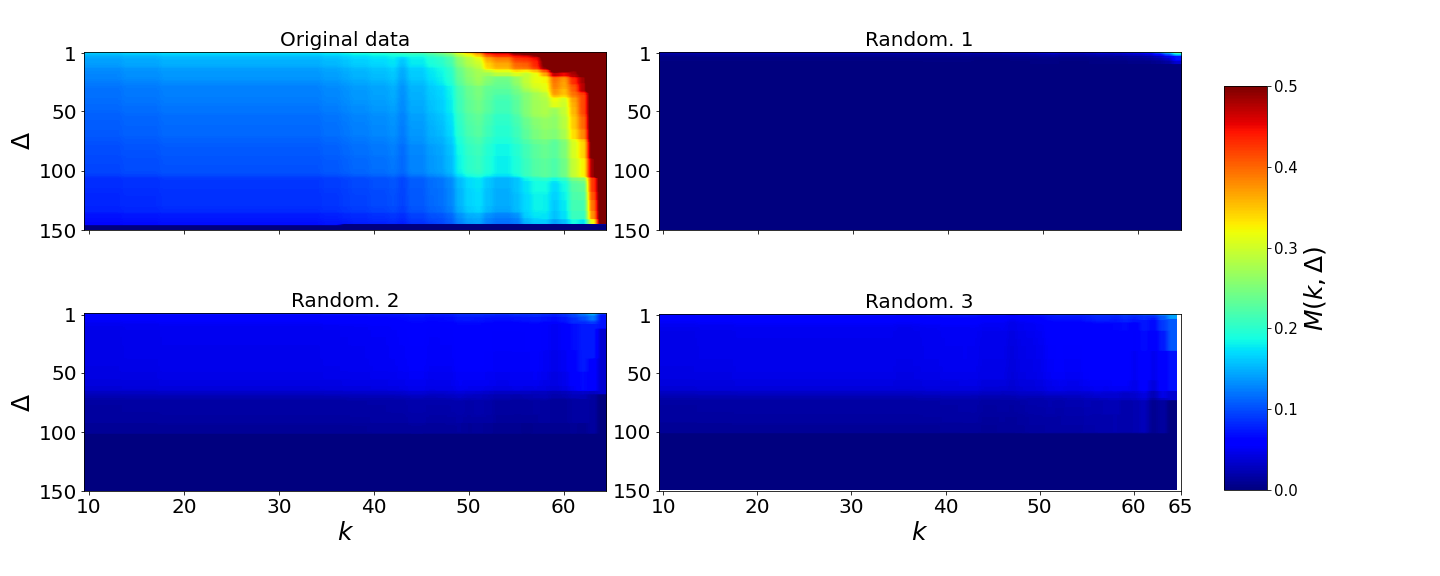}
    \caption{\textbf{Neuronal assembly:} the four plots correspond to the $k-\Delta$ diagrams of the maximal cohesion $M(k,\Delta)$ of the original data set (top left, shown also in Figure 4.a), the \emph{Timestamps reshuffling} (top right) randomization of the temporal network and the \emph{Event} (bottom left) and \emph{Degree-constrained link} (bottom right) shufflings as described in Methods.}
    \label{fig:SI_5}
\end{figure}

\begin{figure}[h!]
    \centering
    \includegraphics[width=\linewidth]{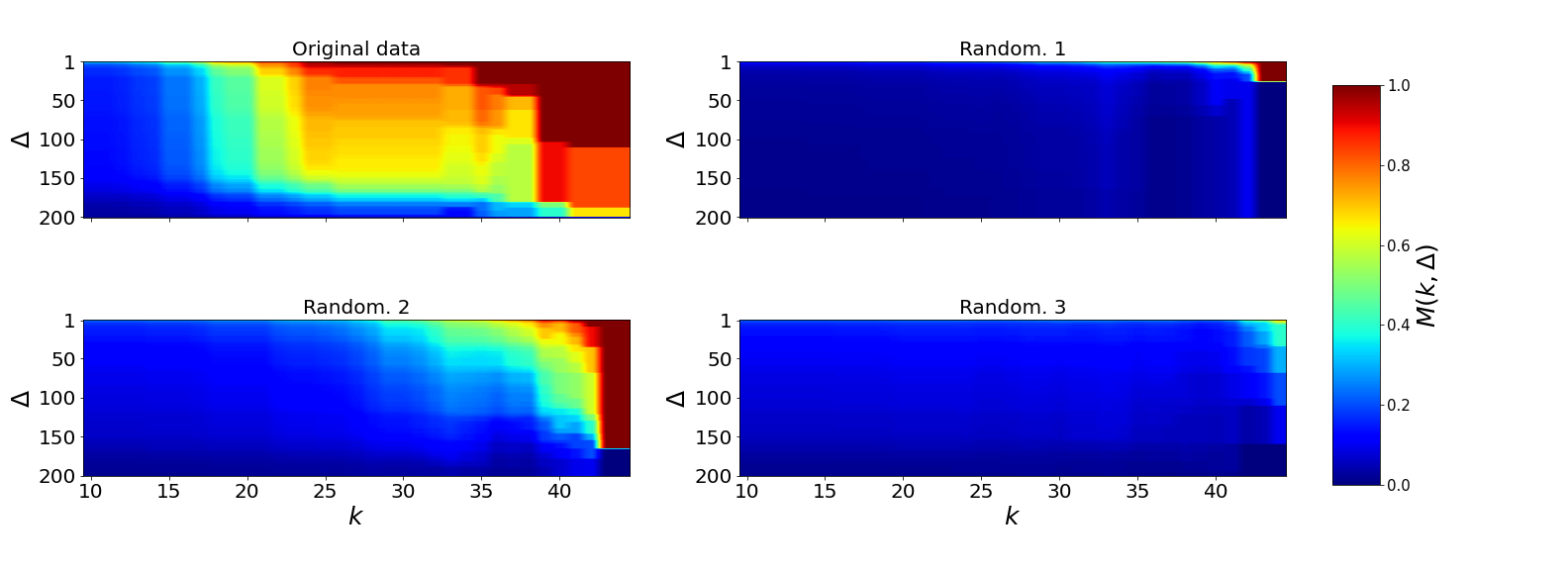}
    \caption{\textbf{Network state $1$ of neuronal assembly:} the four plots correspond to the $k-\Delta$ diagrams of the maximal cohesion $M(k,\Delta)$ computed within network state $1$ of the original data set (top left, shown also in Figure 4.f), the \emph{Timestamps reshuffling} (top right) randomization of the state-wise temporal network and the \emph{Event} (bottom left) and \emph{Degree-constrained link} (bottom right) shufflings as described in Methods.}
    \label{fig:SI_6}
\end{figure}

\begin{figure}[h!]
    \centering
    \includegraphics[width=\linewidth]{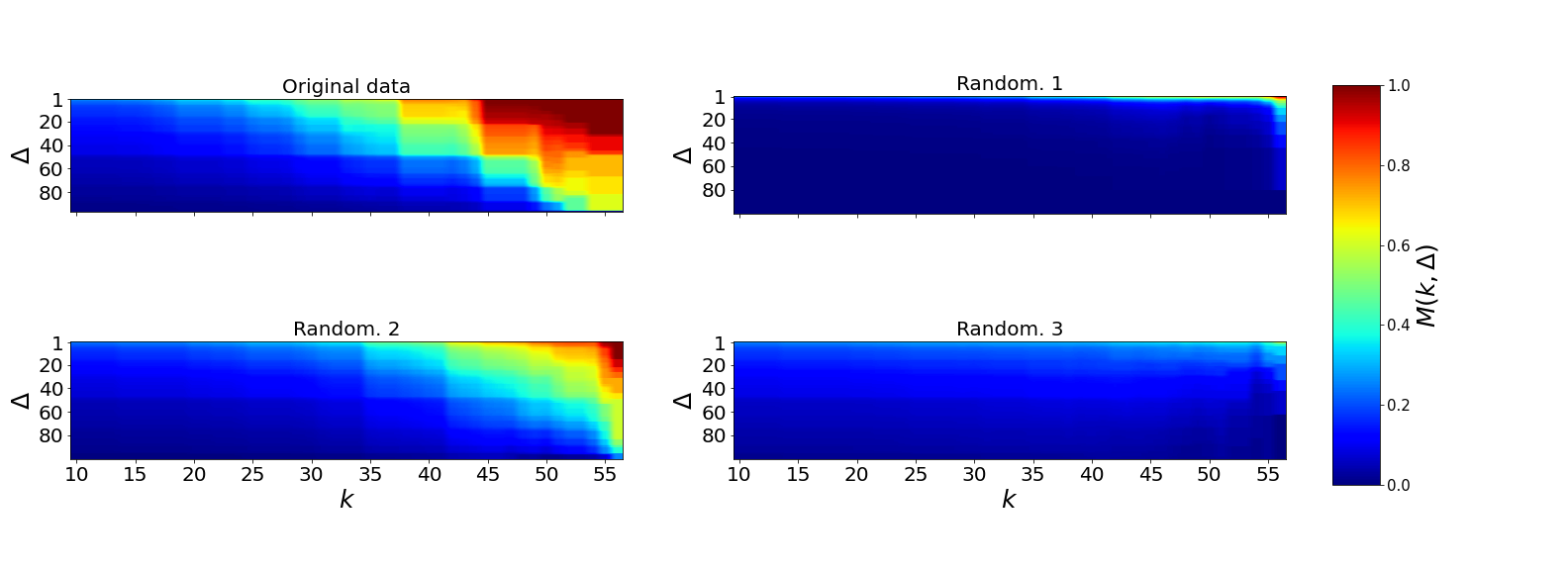}
    \caption{\textbf{Network state $3$ of neuronal assembly:} the four plots correspond to the $k-\Delta$ diagrams of the maximal cohesion $M(k,\Delta)$ computed within network state $3$ of the original data set (top left, shown also in Figure 4.g), the \emph{Timestamps reshuffling} (top right) randomization of the state-wise temporal network and the \emph{Event} (bottom left) and \emph{Degree-constrained link} (bottom right) shufflings as described in Methods.}
    \label{fig:SI_7}
\end{figure}

\begin{figure}[h!]
    \centering
    \includegraphics[width=\linewidth]{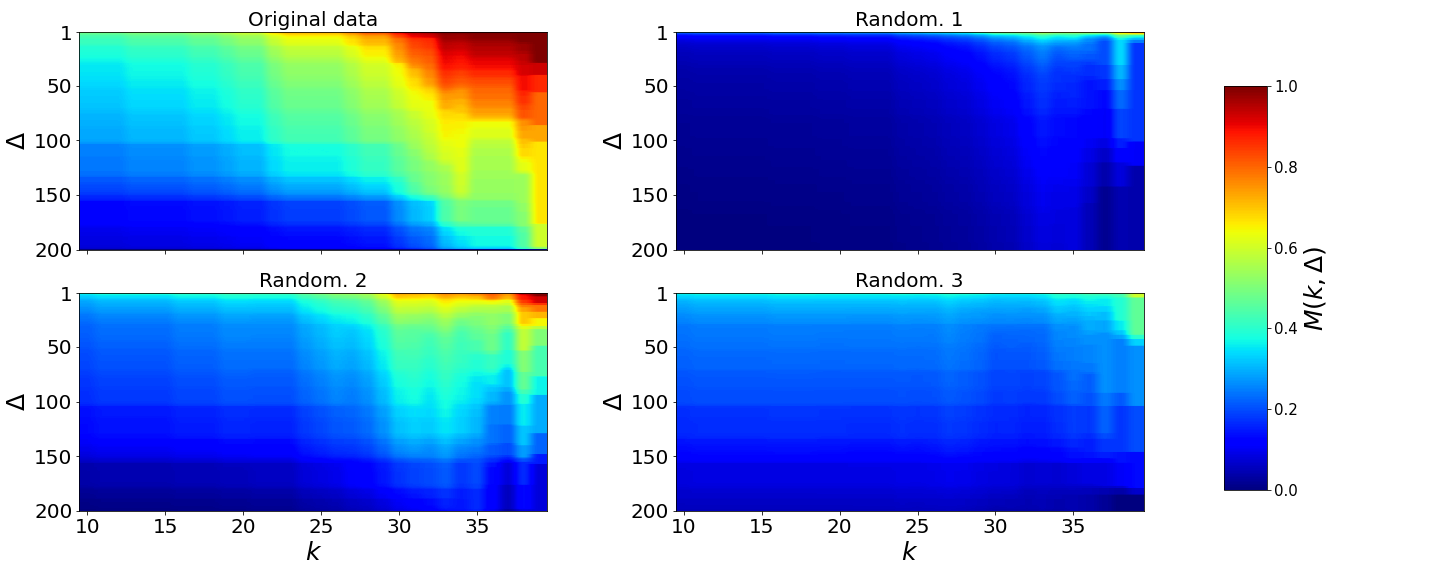}
    \caption{\textbf{Network state $5$ of neuronal assembly:} the four plots correspond to the $k-\Delta$ diagrams of the maximal cohesion $M(k,\Delta)$ computed within network state $5$ of the original data set (top left, shown also in Figure 4.h), the \emph{Timestamps reshuffling} (top right) randomization of the state-wise temporal network and the \emph{Event} (bottom left) and \emph{Degree-constrained link} (bottom right) shufflings as described in Methods.}
    \label{fig:SI_8}
\end{figure}

\end{document}